\documentclass[sigconf]{acmart}
\usepackage{algorithm}
\usepackage{algorithmic}
\usepackage{makecell}
\usepackage{multirow}
\usepackage{textcomp}
\usepackage{enumerate}

\copyrightyear{2024}
\acmYear{2024}
\setcopyright{acmlicensed}\acmConference[IVA '24]{ACM International Conference on Intelligent Virtual Agents}{September 16--19, 2024}{GLASGOW, United Kingdom}
\acmBooktitle{ACM International Conference on Intelligent Virtual Agents (IVA '24), September 16--19, 2024, GLASGOW, United Kingdom}
\acmDOI{10.1145/3652988.3673918}
\acmISBN{979-8-4007-0625-7/24/09}

\begin{document}

\title{Focus Agent: LLM-Powered Virtual Focus Group}

\author{Taiyu Zhang}
\email{taiyu.zhang@kuleuven.be}
\orcid{1234-5678-9012}

\affiliation{%
  \institution{KU Leuven}
  \streetaddress{Naamsestraat 22}
  \city{Leuven}
  \country{Belgium}
  \postcode{3001}
}

\author{Xuesong Zhang}
\email{xuesong.zhang@kuleuven.be}

\affiliation{%
  \institution{KU Leuven}
  \streetaddress{Naamsestraat 22}
  \city{Leuven}
  \country{Belgium}
  \postcode{3001}
}

\author{Robbe Cools}
\email{robbe.cools@kuleuven.be} 
\affiliation{%
  \institution{KU Leuven}
  \city{Leuven}
  \country{Belgium}
}

\author{Adalberto L. Simeone}
\email{adalberto.simeone@kuleuven.be} 
\affiliation{%
  \institution{KU Leuven}
  \city{Leuven}
  \country{Belgium}
}


\begin{abstract}
In the domain of Human-Computer Interaction, focus groups represent a widely utilised yet resource-intensive methodology, often demanding the expertise of skilled moderators and meticulous preparatory efforts. This study introduces the ``Focus Agent,'' a Large Language Model (LLM) powered framework that simulates both the focus group (for data collection) and acts as a moderator in a focus group setting with human participants. To assess the data quality derived from the Focus Agent, we ran five focus group sessions with a total of 23 human participants as well as deploying the Focus Agent to simulate these discussions with AI participants. Quantitative analysis indicates that Focus Agent can generate opinions similar to those of human participants. Furthermore, the research exposes some improvements associated with LLMs acting as moderators in focus group discussions that include human participants. 
\end{abstract}




\maketitle
\section{Introduction}
In the domain of qualitative research, focus groups have emerged as a widely adopted methodology and are extensively employed in both industrial and academic contexts~\cite{kitzinger1994methodology, kitzinger1995qualitative, mazza2006evaluating}, thanks to its structured group discussions aimed at gaining in-depth insights into specific issues. Within Human-Computer Interaction (HCI), researchers routinely employ focus groups as a vital tool in project planning, evaluation, and data collection endeavours~\cite{mazza2006evaluating, troshani2021we, selter2023end, stalmeijer2014using}. Particularly noteworthy is the growing prominence of virtual focus groups, especially in the post-COVID-19 era~\cite{keen2022challenge}. This transition towards virtual focus groups can be attributed to their blending a methodologically sound approach with the potential of  engaging with geographically dispersed and otherwise challenging to access populations~\cite{turney2005virtual}.

Organising a focus group presents two primary challenges: first, gathering so many people at the same time is not an easy task, especially when researchers are interested in exploring the lived experiences of diverse or hard to reach groups~\cite{bruggen2009critical, gratton2011communication, wirtz2019computer}; second, the success of a focus group relies on an experienced moderator with domain-specific expertise. A moderator lacking experience can disrupt the discussion flow or gather unproductive data~\cite{nagle2013methodology}. These issues have sometimes hindered the adoption of focus groups into certain HCI research efforts~\cite{rosenbaum2002focus}.

\begin{figure} [!tbp]
  \centering
  \includegraphics[scale=0.4]{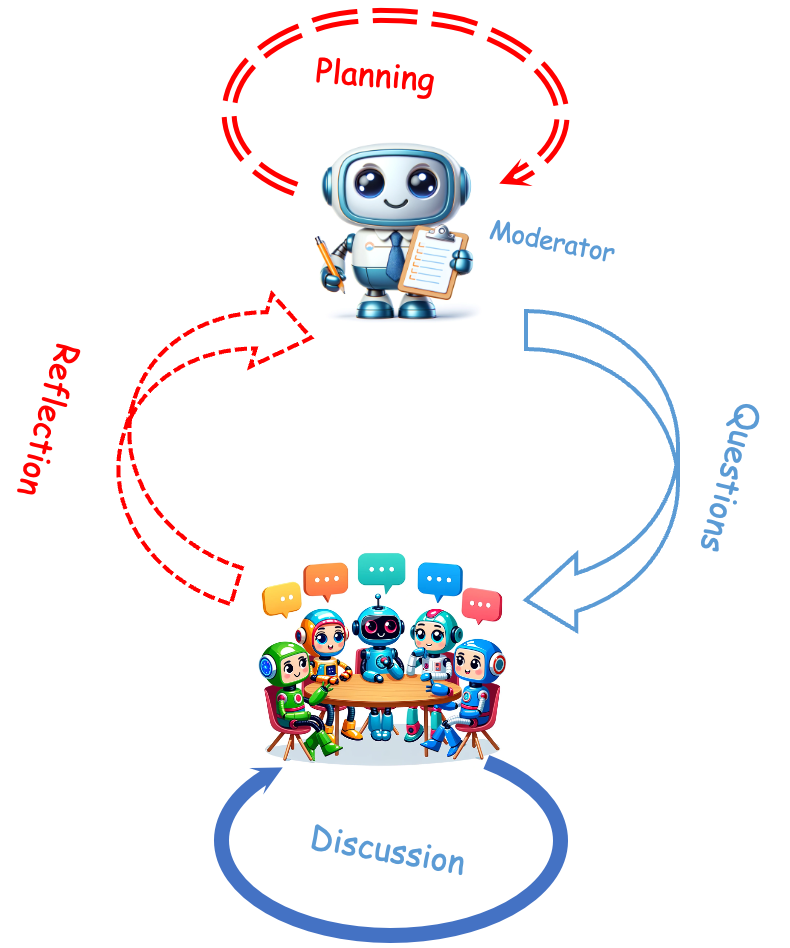}
  \caption{The AI moderator generates questions according to the discussion content and plan, while AI Participants discuss the prompt from the moderator.}
  \label{FocusGroupSimulation}
\end{figure}

The advent of Large Language Models (LLMs), such as ChatGPT, offers a potential solution. These models can frequently communicate in text, generate diverse content from various perspectives based on the large scale of text information on the internet~\cite{reynolds2021prompt, brown2020language}, and demonstrate expertise across several fields, including social sciences, healthcare, and education~\cite{koubaa2023gpt, sallam2023chatgpt}. Their capabilities extend to assisting with paper writing~\cite{katar2023evaluation, ciaccio2023use}, providing legal advice~\cite{katz2024gpt, nay2024large}, and supporting medical inquiries~\cite{haupt2023ai}. Given these advancements, focus groups, a classic qualitative data collection method, should benefit from LLMs. Despite their potential, these models are prone to certain limitations such as misunderstanding human instructions, generating potentially biased content, or factually incorrect (hallucinated) information~\cite{wang2023aligning}. Additional framework design is still necessary for multi-agent tasks, such as societal simulations~\cite{park2023generative} or role-playing game simulations~\cite{xu2023exploring}.

This work introduces the ``Focus Agent'', an LLM-based moderator for focus groups that has two functions: 1) simulating discussions without human participants and collecting AI-generated opinions, and 2) guiding focus groups as a moderator as shown in \autoref{FocusGroupSimulation}, with human participants as well. To address prevalent issues in multi-agent simulations, including repetitive opinions and the generation of irrelevant content, the ``Focus Agent'' employs a scheduled discussion format that divides the focus group into distinct stages, each corresponding to a specific topic. This method mirrors the strategies employed by experienced human moderators. Additionally, the framework incorporates reflection periods during discussion to counteract memory loss during the simulation, ensuring a coherent and productive discussion flow. When moderating focus groups with human participants, a multi-person Speech-to-Text (S2T) and Text-to-Speech (T2S) integration enables the ``Focus Agent'' to interact with multiple users simultaneously.

Our work primarily explores the application of LLMs in simulating focus group discussions. Two main Research Questions (RQs) are as follows:

\textit{ RQ 1: To what extent do the opinions generated by a LLM align with those of human participants in focus group?}

\textit{ RQ 2: To what extent is a LLM effective in performing the duties of a moderator in focus group discussions?}

To answer these RQs, we conducted a user study with 23 participants across five discussion groups. Participants engaged in a one-hour AI-moderated focus group discussion on the topic of ``digital well-being'', followed by a 30-minute session led by a researcher to share their experiences, evaluate the AI moderator's performance and collect feedback, which was referred as a \textit{meta focus group} in our work. Meanwhile, the Focus Agent simulated the focus group discussions on the same topic with AI participants. Qualitative analysis including thematic analysis and content analysis of the transcriptions reveals that the AI simulation outputs the majority of opinions expressed by human participants. Additionally, we assessed the performance of the Focus Agent functioning as a moderator, both in the focus group simulation with AI participants as well as with focus groups involving human participants. Based on our findings, the Focus Agent meets the essential criteria required of a focus group moderator. This includes progressively guiding discussions from general to more specific topics and maintaining an actively engaged atmosphere, drawing on the fundamental literacy expected of a focus group moderator~\cite{stewart2014focus}. However, when tasked with moderating discussions involving human participants, the agent's ability to interact with humans seems constrained, and it has not demonstrated sufficient understanding of human conversation. We identified several limitations of current LLMs in managing multi-person discussions and offer suggestions for integrating AI agents into focus group more effectively. To promote further research, the code has been open-sourced\footnote{\url{https://github.com/AriaXR/FocusAgent}}.

\section{Related Work}
This section discusses previous research directly related to our study. We divided it into three subsections: Focus Group Development, Multi-Agent Simulation and Multi-speaker speech recognition for Voice-based Conversational Agents.

\subsection{Focus Group Development}

The utilisation of focus groups, or group depth interviews, is a cornerstone method within the realms of advertising, marketing, and HCI research due to its effectiveness in gathering qualitative insights~\cite{stewart2014focus}. The earliest focus groups were conducted through face-to-face conversations, which make the organisation complex and time-consuming, even with a lot of fees for participant reimbursement~\cite{rosenbaum2002focus}. The popularity of online focus groups has augmented their appeal, offering advantages such as the convenience of participation from any location at any time, and anonymity, which reduces participants' apprehension of judgement~\cite{daniels2019steer, wilkerson2014recommendations, stewart2017online}. Researchers inviting many people to participate in online meetings at the same time often encounter difficulties, such as inconsistent time schedules, time differences, and poor communication caused by network delays. To further facilitate users' participation in focus groups, some social media platforms provide asynchronous text-based focus groups~\cite{gordon2021asynchronous, biedermann2018use, richard2021guide, wenzek2019ccnet}. However, as participants do not contribute simultaneously, it brings some difficulties relating to such a reduced `spontaneity'~\cite{bruggen2009critical, nicholas2010contrasting} including: shorter answers with fewer word counts~\cite{chen2019texting}; uneven flow during the interactions due to their lag~\cite{veloso2020whatsapp}; and more unfocused exchanges that do not always address the relevant research question~\cite{bruggen2009critical}. 

The recent advancements in LLMs, which are trained on extensive internet text data, offer novel opportunities for conducting focus groups. As an innovative retrieval model, LLMs have the potential to streamline the data collection process~\cite{zhu2023large}. Utilising LLMs to simulate focus groups presents a simpler and potentially more efficient alternative to engaging human participants, thereby opening new avenues for qualitative research.

\subsection{Multi-Agent Simulation}

Despite the capability of LLMs to process one-on-one question-answer formats, their deployment in long term dialogues and opinion generation, such as focus group discussions, reveals some limitations. These challenges include difficulties in understanding complex instructions, hallucination of agents, a limited token memory leading to loss of continuity, repetitive dialogues, and the generation of meaningless conversation in long-term interactions~\cite{openai2023gpt, xu2024hallucination}. 

To help solve these issues, recent research has come up with new ways to organise how these AI agents think and respond, tailored to specific kinds of tasks~\cite{talebirad2023multi, park2023generative}. 
The Chain-of-Thought (CoT) principle is pivotal, serving as the foundational idea behind them~\cite{wang2023survey}. By dissecting complex issues into simpler elements, it facilitates a collaborative approach among multiple agents to tackle each component, leading to a comprehensive solution. By decomposing complex problems into many simple parts, the solution is achieved through the combined efforts of multiple small agents. Additionally, the reflection mechanism plays a crucial role in addressing memory limitations and enhancing the authenticity of the generated content~\cite{yan2024mirror}. This process involves storing detailed historical data as structured information, which can be referenced for more informed decision-making in future interactions. Moreover, to improve the consistency of agent performance across various contexts, some works have investigated the exploration of diverse prompting techniques tailored to the specific roles~\cite{shanahan2023role}.

In our work, we have built upon insights from previous research to address potential challenges that could arise during focus group discussions. Furthermore, we have developed a novel framework for conducting focus groups, primarily guided by an AI moderator. The AI moderator facilitates simulated focus group discussions and aids in coordinating focus groups that include human participants. To bridge the interaction gap with human participants, we incorporate a voice-based conversational agent to the moderator.

\subsection{Multi-speaker speech recognition for Voice-based Conversational Agents}

Unlike text-based chatbots, Voice-based Conversational Agents (VCAs) necessitate an extra technological layer for operation: they use a speech-to-text (S2T) process to interpret spoken inputs and a text-to-speech (T2S) system for generating spoken responses~\cite{jokinen2022spoken, rough2020don}. This integration allows VCAs to facilitate interactions in a more natural, conversational manner, bridging the gap between human users and digital assistants. 

However, current S2T technologies, such as Google's API or OpenAI's Whisper, encounter difficulties in long-term group discussions such as focus groups~\cite{radford2022robust}. One challenge with using S2T technologies like Whisper for multi-participant discussions is the duration limit on voice recording inputs, which is considerably less than the typical length of conversations. A potential solution involves segmenting longer discussions into shorter fragments using Voice Activity Detection (VAD), which helps manage recordings more effectively~\cite{bain2022whisperx}. Another limitation is lack of speaker differentiation, a critical feature for understanding who is speaking in group discussions. Some research has attempted to identify individual speakers by analysing the unique timbre of their voices~\cite{medennikov2020target, horiguchi2021end,horiguchi2020end}. However, these methods often fall short in accuracy due to the absence of prior information about the speakers. A more effective approach involves using a pre-recorded sample from each speaker, enabling a retrieval-based method to significantly improve performance by accurately distinguishing between speakers~\cite{desplanques2020ecapa}.

In our work, we improved Whisper, an open-source S2T model, with a retrieval-based technique, optimising it for multi-participant discussions such as focus groups.

\section{Focus Agent Implementation}

Our Focus Agent was designed to simulate focus group discussions and facilitate running sessions involving human participants. For the focus group simulation, we devised a multi-agent framework, complemented by a moderator to oversee the entire focus group process. This ensures that the contributions from AI participants are both relevant and valuable. Regarding interactions with actual human participants, we incorporated S2T and T2S systems into the AI moderator, enabling voice-based communication. 

\subsection{Focus Group Simulation}

\begin{figure}[!thbp]
    \centering
\includegraphics[scale=0.14]{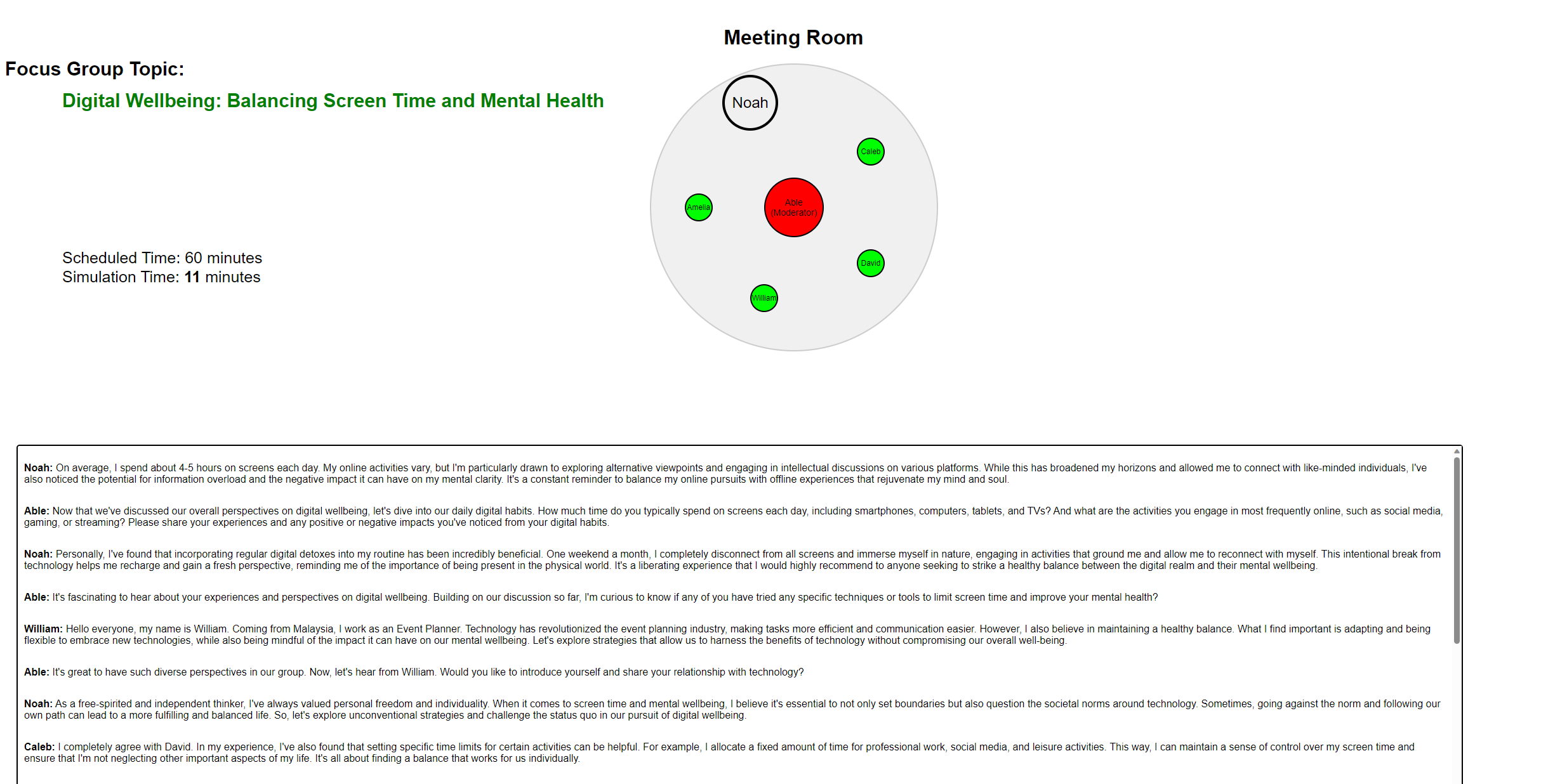}
\caption{A web demo of the Focus Group simulation system.}
\label{simulation}
\end{figure}

In accordance with the benchmark study conducted by OpenCompass~\cite{2023opencompass}, the two most advanced Large Language Models (LLMs) available in the field at the time of writing are ChatGPT and GPT4. Pilot testing revealed that ChatGPT resulted in similar opinions compared with GPT4, after which we decided not to use the superior GPT4 due to its 20-fold increase in cost. Compared to direct prompts, our algorithmic framework improves the realism and comprehensiveness of the AI simulation, as corroborated in \autoref{simulation}.

Initially, we attempted to employ a singular prompt to simulate focus group discussions. However, concerning both content and length, the generated outcomes significantly deviated from our expectations. In response to these challenges, we introduce the framework of our Focus Agent, featuring an AI moderator to guide the discussion process. As shown in \autoref{FocusGroupSimulation}, this AI moderator generates some plans to divide the whole discussion into multiple stages, aligning with the distinct topic and aims of the focus group. Based on these guidelines, the AI moderator then facilitates a simulated focus group discussion with other AI entities as participants. Throughout the conversation, the moderator actively engages in reflection, responding to the dialogue of the participants by timely introducing pertinent questions to foster further discussion. We explained this process in detail in the online appendix.

Within the simulated focus group, each participant represents an artificial intelligence entity. Experimenters are responsible for defining key parameters such as the topic, goals, overall duration, and specific characteristics of the participants, which include names, ages, occupations, nationalities, and personalities. In this setting, LLMs are tasked with understanding the context through assigned roles, typically categorised as system, user, and assistant. The system role involves attributing virtual personas to the LLMs, while the user and assistant roles are designed to aid in interpreting the context either from the viewpoint of the designated character or from that of others. To achieve this, we have developed a sequence of prompt designs, the details of which are provided in the online appendix.

To simulate the focus group discussion as realistically as possible, we designed the algorithm of both moderator and participants. The role of the moderator within the focus group simulation system encompasses the critical responsibilities of guiding and orchestrating the discussion, which includes managing time allocation and steering the discourse topics. These responsibilities are reflected in the moderator's thought chain, elucidated in Algorithm \autoref{Moderator}. We added a reflection mechanism at the end of every stage to compress the context of previous discussion to avoid memory lost. Time allocation is managed based on text lengths, with a convention of one hundred words equating to approximately one minute within the simulation.

\begin{algorithm}
    \caption{Moderator}
    \label{Moderator}
    \renewcommand{\algorithmicrequire}{\textbf{Initialisation:}}
    \renewcommand{\algorithmicensure}{\textbf{Output:}}
    \begin{algorithmic}
        \REQUIRE $List:[Stages], List:[Time Arrangements]$
        \ENSURE $Str: Response$
        \FORALL{$stage,time_{stage} \leftarrow Stages,Time Arrangements$}
            \STATE $Response \leftarrow LLM(NewStagePrompt)$
            \STATE $time_{cur} \leftarrow Estimate(Response)$
            \WHILE{$time_{cur} < time_{stage}$}
            \IF{Response from participants}
            \STATE $Response  \leftarrow ParticipantResponse$
            \ELSIF{any participant is inactivate}
            \STATE $Response  \leftarrow LLM(InactivateParticipantPrompt)$
            \ELSE
            \STATE $Response \leftarrow LLM(InsightsPrompt)$
            \ENDIF
            \STATE Update $time_{cur}$ according to $Estimate(Response)$ 
            \ENDWHILE
        \ENDFOR
    \end{algorithmic}
\end{algorithm}

Algorithm \autoref{Panticipants} outlines the systematic approach adopted by each AI participant throughout the discussion, with their level of engagement assessed by the LLM. The LLM dynamically evaluates the ongoing conversation and the contributions of other AI participants to gauge engagement levels. AI participants are provided the latitude to contribute to the discussion uninterrupted unless they surpass the stipulated time allocation. In instances where participants opt to disengage or exhibit novel ideas, signalling a lull in the discourse, the moderator intervenes by posing new questions, drawing inspiration from the preceding discussions. In parallel, the moderator actively encourages less active participants to actively partake in the discourse. Participant activity is monitored through the detection of speaking times within the ongoing stage. Participants who exhibit negligible speaking activity or speaking three times less than those of the most speaking participants are categorised as inactive.

\begin{algorithm}
    \caption{Participants}
    \label{Panticipants}
    \renewcommand{\algorithmicrequire}{\textbf{Initialization:}}
    \renewcommand{\algorithmicensure}{\textbf{Output:}}

    \begin{algorithmic}
        \REQUIRE $List:[Participants]$
        \ENSURE $Str: response$
        \REPEAT
        \STATE $engagements \leftarrow []$
        \FORALL{$participant$ \textbf{in} $Participants$}
        \STATE $engagements$ add $LLM(EngagementPrompt)$
        \ENDFOR
        \IF{$Max(engagements) \geq Threshold$}
        \STATE $speaker \leftarrow Participants[Index(Max(engagements))]$
        \STATE $Response \leftarrow LLM(PartResponsePrompt(speaker))$
        \ENDIF
        \UNTIL{Finished}
    \end{algorithmic}
\end{algorithm}

\subsection{Voice-based Focus Agent with human participants}

\begin{figure}[!thbp]
  \includegraphics[scale=0.25]{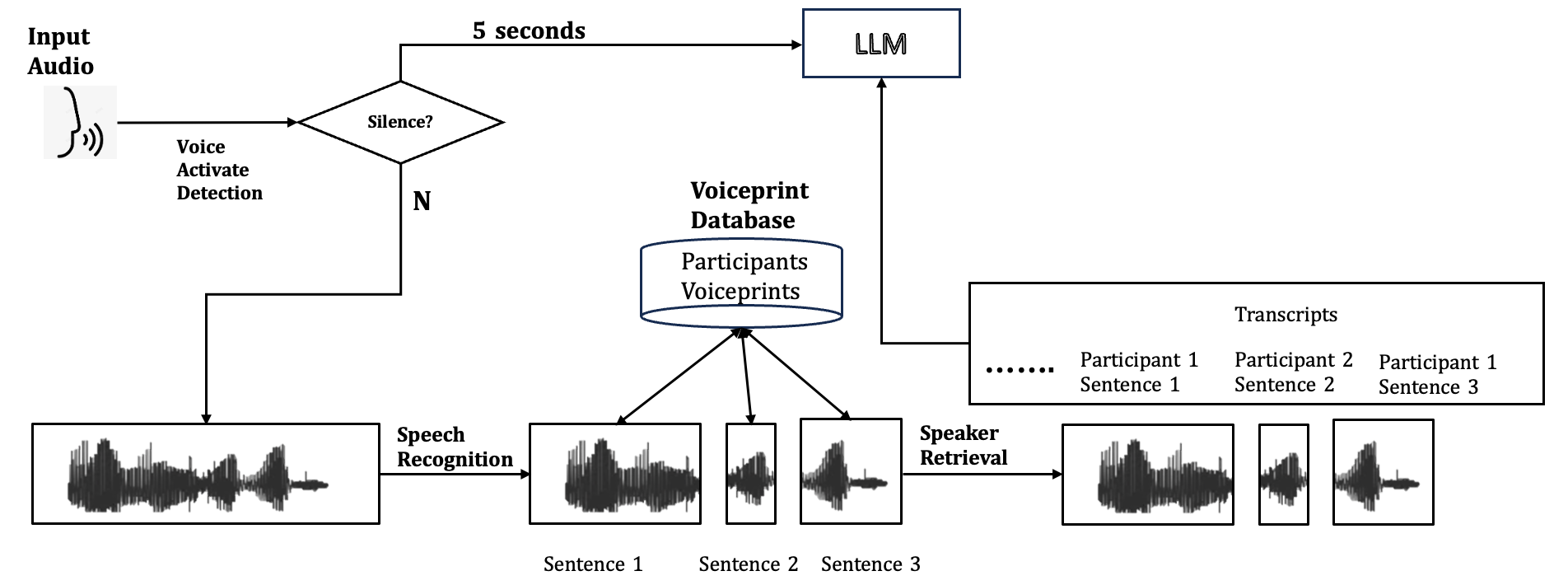}
  \caption{Speech to Text system. We divided long audio recording into short pieces with voice activity detection. Then we transcribed the short audio pieces and recognised the speaker according to the voiceprints collected in advance from the participants.}
  \label{S2T}
\end{figure}

To make sure the AI moderator can communicate with human participants efficiently, S2T and T2S are necessary. APIs provided by various companies are often suitable for many scenarios. However, they fall short of our specific needs for facilitating multi-participant discussions in focus groups due to limitations related to the length of input recordings and the absence of speaker differentiation. To address these challenges, we have developed our own S2T system, as depicted in \autoref{S2T}. This system processes long discussion audio by segmenting it into shorter sentence-length audio pieces, leveraging VAD for segmentation. Subsequently, it identifies the most similar participant from a database of participant voices and transcribes the audio using the open-source S2T model Whisper by OpenAI~\cite{radford2022robust}. To ensure participants have ample opportunity to express their views without undue interruption, the AI moderator is programmed to intervene only after a silence of 5 seconds, thus differing from approaches that might actively disrupt the conversation flow.

In order to incorporate T2S functionality into our system, we leveraged the Google TTS API\footnote{https://console.cloud.google.com/speech/text-to-speech}. To allow some participants who are interested in discussion in an immersive environment, we established the focus group environment within Mozilla Hubs\footnote{https://hubs.mozilla.com}, a Virtual Reality (VR) platform.

\section{Pilot Study}

To enhance user experience, we conducted a pilot study with four volunteers before the main user study to assess the system’s stability and the AI moderator’s effectiveness. The pilot included a 50-minute focus group discussion and a 30-minute feedback session on the AI agent’s performance.

Feedback from the pilot highlighted areas for improvement, which were addressed to optimise the user study:

\begin{enumerate}
    \item[1.] Human participants may not always have insights for every query, unlike AI participants who consistently generate new content. Observations showed the AI moderator might repeat questions if there were no responses, leading to stagnation. We adjusted the AI moderator’s protocol to move on if no further responses were forthcoming.
    \item[2.] Anonymity in Summaries: Volunteers were uncomfortable with being mentioned by name in summaries. We revised the process to ensure participant anonymity, enhancing comfort levels.
    \item[3.] Conciseness of Questions: The long content generated by LLMs are not ideal for verbal interactions. We refined prompts to yield shorter responses.
\end{enumerate}
 
Additionally, we assessed the S2T system’s accuracy to ensure comprehensive transcription and understanding by the agent. The Word Error Rate (WER)\footnote{WER is a metric for gauging speech-to-text conversion accuracy, calculated as $WER=(S+D+I)/N$, where $S$ denotes substitutions, $D$ deletions, $I$ insertions, and $N$ the total number of words in the reference text.} served as the evaluation metric. Professional human transcribers typically achieve a WER of 11.3\% in open conversational settings~\cite{xiong2016achieving}. Using this as a benchmark, we found our S2T system achieved a WER of 4.6\%, demonstrating commendable accuracy. For speaker identification, our system achieved a micro F1 Score of 0.81 using the EN2001 audio segment from the AMI Corpus~\cite{carletta2006announcing}, highlighting its capability in recognising speakers. The pilot study indicated the agent exhibited no significant misunderstandings of the conversations.

\section{User Study}

\begin{figure}[H]
    \centering
\includegraphics[width=\linewidth]{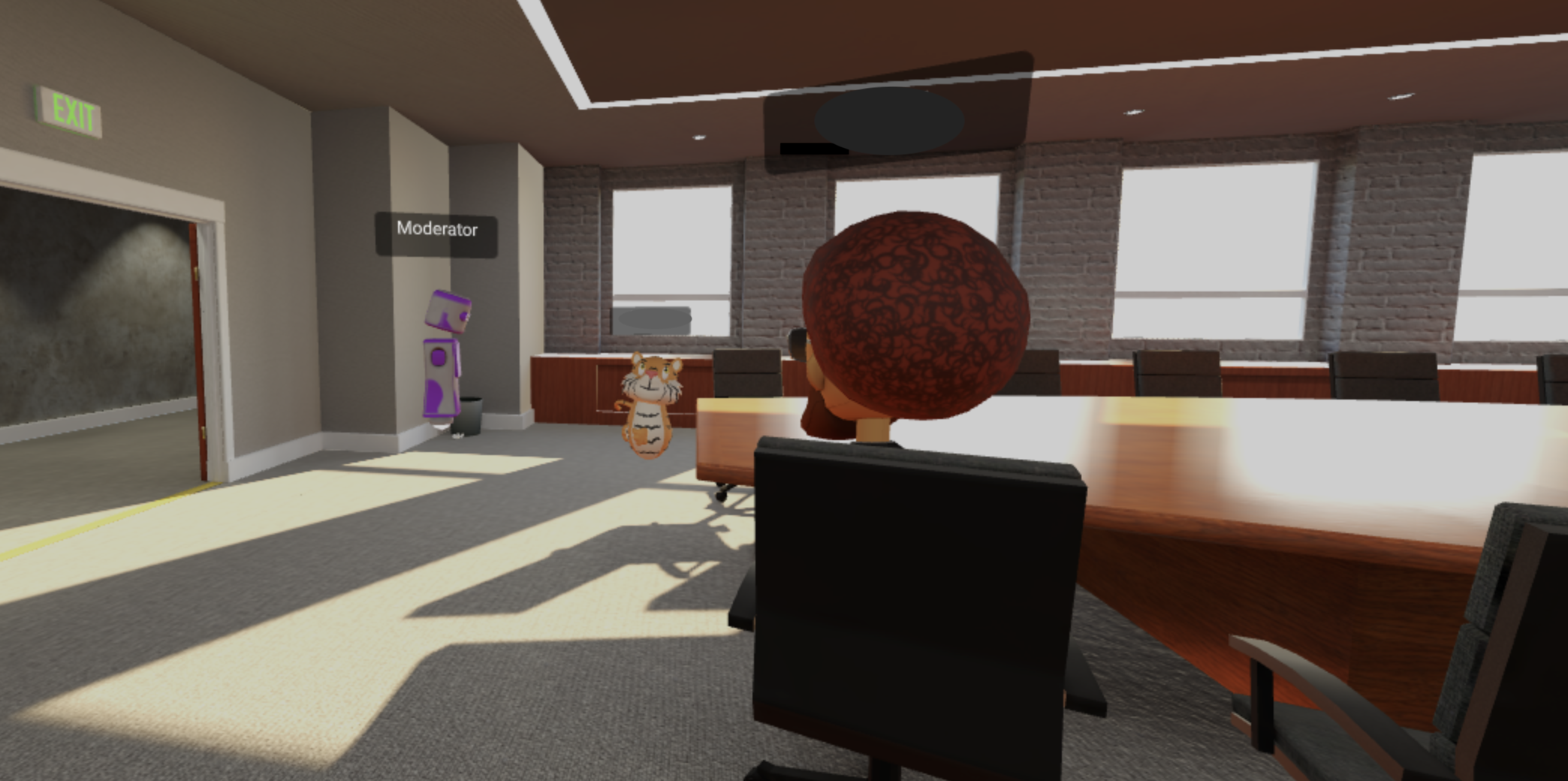}
\caption{Users participant focus group using Focus Agent in VR environment.}
\label{FocusAgent}
\end{figure}

To investigate our research questions, we designed a user study that involved human participants engaging in focus group discussions on the theme of ``digital well-being,'' alongside simulations of focus groups centred around the same topic. The objective of these sessions was to study individual practices in managing screen time and their perceptions of its impact on mental health. The choice of ``digital well-being'' as the focal topic was strategic, given its universal relevance, which facilitated participant recruitment. Participants had the option to join the focus groups either via a VR headset or through their personal computers, aiming for device consistency within groups to streamline the discussion dynamics, as shown in \autoref{FocusAgent}. 

\textit{Demographics.} Our recruitment efforts yielded 23 participants, where we assigned 11 to join with VR headset and 12 to join with their own personal computer. The participant pool had an average age of 30 years ($min=18$, $max=60$, $SD=10$), distributed across five groups--three with VR headset and two with desktop. Each group comprised 3 to 6 individuals, ensuring a diverse range of perspectives and experiences. The selection of the total number of groups is based on previous work~\cite{guest2017many}, which has demonstrated that five groups are optimal for focus group studies.

\textit{Procedure.} The user study included three distinct components: a primary focus group involving human participants (hereafter referred to as ``\textit{focus group}''), a meta focus group where human participants convened to reflect on their experiences within the \textit{focus group} (hereafter referred to as ``\textit{meta focus group}''), and a simulated focus group with AI entities as participants (hereafter referred to as ``\textit{focus group simulation}'').

First, participants submitted a one-minute self-introduction audio recording before the focus group. This recording collected demographic information (age, prior focus group experience, and daily screen usage) and provided a unique voice print for each participant. This data initialised the AI participants in the simulation. We assessed English proficiency based on the accuracy of the S2T results from their recordings. Then participants accessed the designated meeting rooms in Mozilla Hubs. For VR groups, our team provided VR headsets (Quest series or Vive Pro), while the desktop group used their own PCs. Once all participants were ready, the researcher started the system, and the AI moderator began moderating the focus group. An author observed and recorded essential information throughout the sessions. The sessions were scheduled for 60 minutes, with an actual average duration of 51 minutes ($SD = 13 minutes$).

Following the conclusion of each focus group discussion, a meta focus group was conducted. This session spanned approximately 20 minutes and was facilitated by one of the authors. The topic of the meta focus group mainly focuses on two points: the experience of focus group discussion and the attitude to the AI moderator.

At the end, each participant received a 10€ gift card as compensation. This study was reviewed and approved by the university's ethics review board.

\section{Result Analysis}

Following the methodological framework proposed by \citeauthor{gerling2020virtual}~\cite{gerling2020virtual}, we employed both thematic and content analyses to scrutinise the transcripts derived from the focus group and focus group simulation sessions. Additionally, thematic analysis was specifically applied to the meta focus group discussions to collect participant feedback. For the transcription of data from the user study, we utilised the outputs from our S2T system, subsequently refining these transcripts against the recorded audio by two researchers. The final evaluation of our S2T system showcased a WER of 2.5\% and an F1 score of 0.9, indicating a level of performance sufficiently reliable for the purposes of our study. Due to recording issues, the data from the third focus group session was incomplete. The transcription for this group was reconstructed based on recollections and notes taken by an observer, and consequently, this data was not included in the accuracy assessment of the S2T system. The initial analysis was conducted by the lead author, with the findings subsequently reviewed and validated by the co-authors.

\subsection{Focus Group}

In the thematic analysis conducted on the transcriptions from both the human focus group and the focus group simulations, we elicited distinct themes related to our study topic. From the transcriptions, we identified four central themes. In contrast, the focus group simulations revealed five themes, incorporating an additional theme focused on the challenges associated with controlling screen time. This discrepancy mainly came from the differences in moderation performance between the two groups. In the focus group simulations, the AI moderator tends to guide AI participants to engage more deeply with the topics. While human participants in the focus group did broach additional topics, these were less related to the central theme of discussion, highlighting a contrast in how thematic expansion was handled across the two settings.

\begin{figure*}[!thbp]
\centering
\includegraphics[scale=0.3]{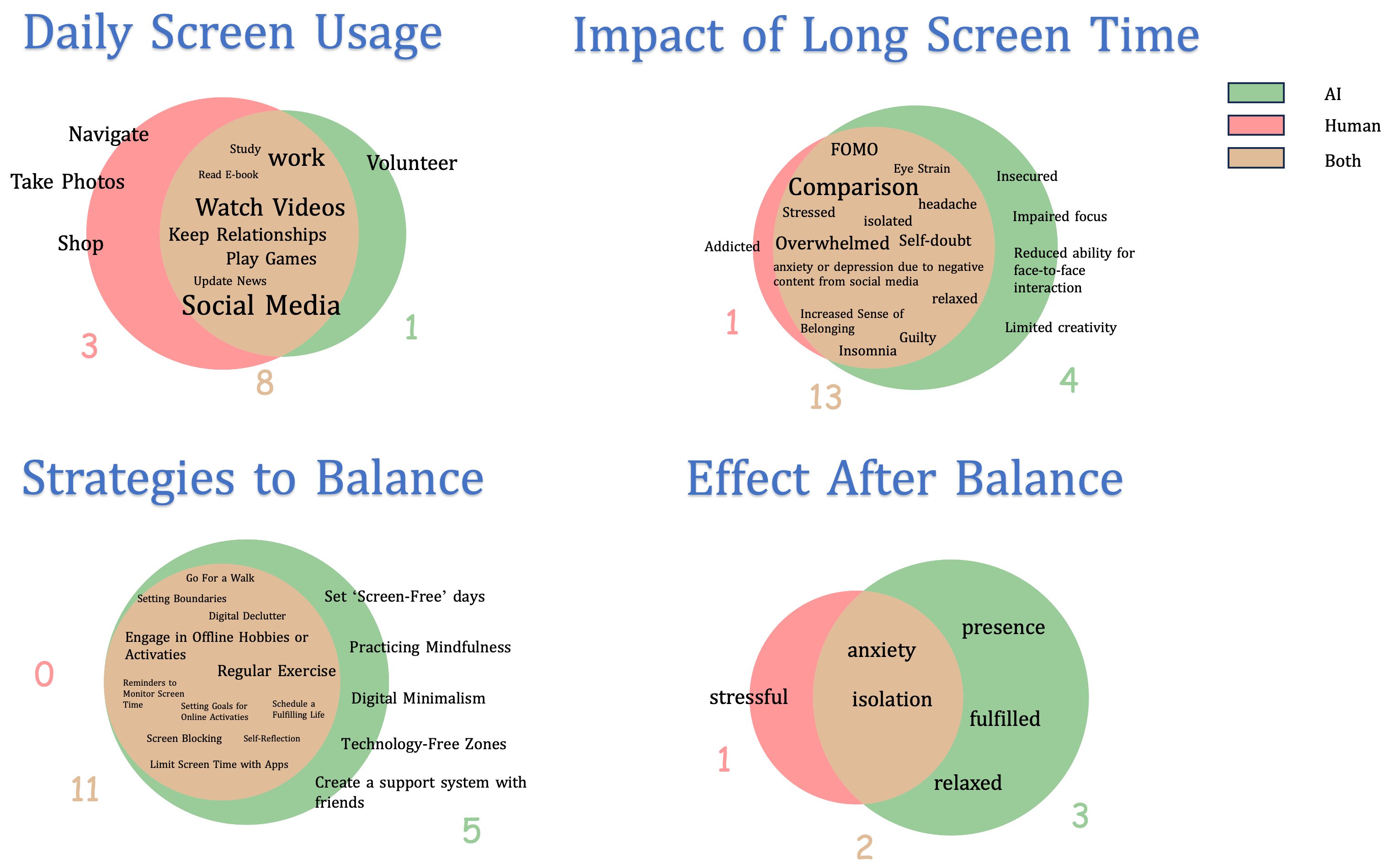}
\caption{Content analysis according to the themes from both focus group and focus group simulation, font size indicates the frequency of the codes}
\label{content_analysis}
\end{figure*}

\begin{figure}[!thbp]
\centering
\includegraphics[scale=0.3]{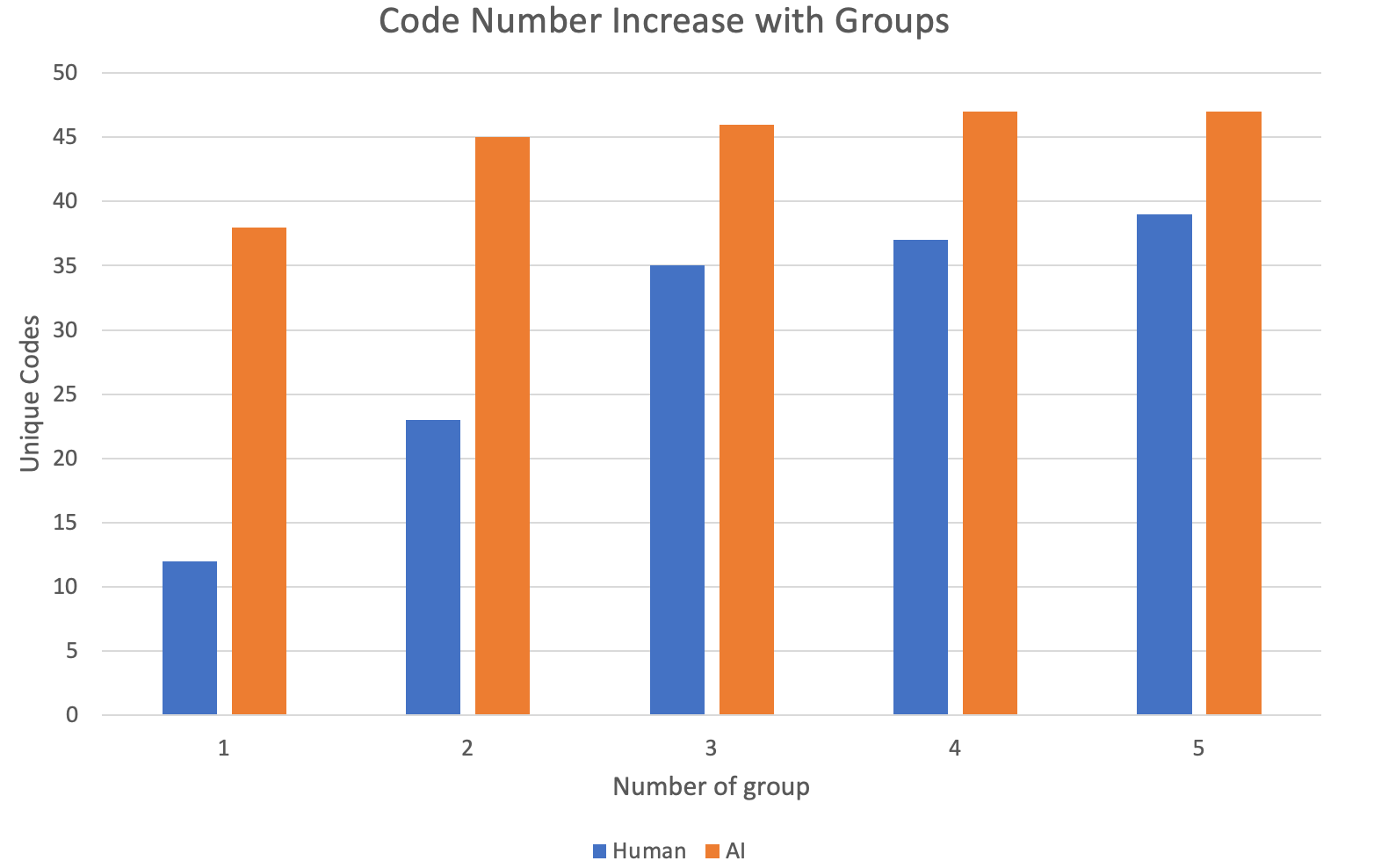}
\caption{Unique code number increased according to the round of focus group and focus group simulation.}
\label{codenumber}
\end{figure}

In our content analysis, we derived 39 unique codes from the focus group transcriptions and 47 from the focus group simulations, each reflecting various facets of the discussion topic. To compare the perspectives of AI and human participants, we illustrated the overlap and divergence of these codes through a Venn diagram, as showcased in \autoref{content_analysis}. The analysis revealed that the majority of opinions expressed by human participants were also covered by AI participants. Interestingly, AI participants introduced several viewpoints not raised by their human counterparts, such as \textit{volunteering online} during daily screen usage, adding additional dimensions to the discussion. Another observation from this analysis is the tendency of AI participants to express similar opinions more than human participants across different focus group sessions. The data referenced in \autoref{codenumber} reveal a discrepancy in code generation between simulation and focus groups. Each iteration of the focus group can collect similar number of codes. The result indicates that simulations of focus groups tend to generate higher repetition of identical codes. Following several iterations, the aggregate of unique codes converges, suggesting that the most common opinions have been collected. At this point, AI participants can not generate new codes, whereas human participants continue to demonstrate potential for such creativity. This observation underscores the tendency of AI to produce more common opinions, while human participants display greater variance and individuality in their perspectives.

\subsection{Meta Focus Group}
\label{sec:meta_focus_group}
According to the transcriptions of the meta focus group, we coded 51 data points and identified three main themes.

We derived three themes from the data: 1) User Experiences of the Virtual Focus Group; 2) User Attitudes towards the Focus Agent, which is further divided into two sub-themes: a) Positive Attitudes, and b) Negative Attitudes; and 3) Feedback on the Virtual Focus Group System.

\paragraph{Theme 1: User Experiences of Focus Group.}

A majority of participants conveyed satisfaction with the focus group discussions, highlighting several reasons. For many, the topics discussed were directly relevant to their daily lives, adding value to their participation. As one participant explained, ``\textit{I think it's great to discuss these topics because that's what we deal with every day.}'' (G5, P4). Furthermore, participants appreciated the diversity of perspectives present, valuing the opportunity to exchange experiences. An exemplifying statement reads, ``\textit{I think you bring up so many great points. It’s very enriching to hear different perspectives.}'' (G5, P3). At the end of the discussions, the moderator inquired whether participants had any additional opinions on the topic that they had not had the opportunity to express during the session. All participants confirmed that they had no further insights to share, indicating that the discussions had comprehensively covered the topic from their perspectives.

\paragraph{Theme 2: Attitude to the Focus Agent.}

The second theme encapsulates the users' feedback and experiences with Focus Agent. This theme is divided into two categories: positive and negative, to provide a clearer understanding of the users' attitudes towards Focus Agent.

SubTheme 1: positive attitude. A prevalent sentiment among participants was their appreciation for the guidance offered by the Focus Agent, acknowledging its efficacy in steering the discussions. As an example, one participant remarked, ``\textit{The moderator kind of did a good job by posing questions that allowed us to express our thoughts and encouraged other participants to share their sentiments on the topic.}'' (G4, P1). Furthermore, three participants specifically commended the Focus Agent's clear articulation in English, while an additional participant admired the agent's friendly demeanour.

SubTheme 2: Negative attitude. The prevailing sentiment among participants leaned towards dissatisfaction with the Focus Agent's performance. A recurring concern revolved around the repetition of questions, as one participant articulated, ``\textit{I found it somewhat confusing at times since the moderator repeated the questions several times, which we had already discussed}'' (G1, P2). Another noteworthy issue was the perceived lack of intellectual acumen exhibited by the Focus Agent during discussions. For example, one participant expressed, ``\textit{I don't believe it possesses true intelligence, nor does it seem capable of comprehending all the information we've conveyed, let alone guiding us into more profound and coherent discussions}'' (G5, P2). At last, some biases were identified in the discussion, notably in steering participants towards articulating the adverse effects associated with prolonged screen use, ``\textit{When discussing the impact of long screen using time, I felt that the AI moderator tried to demonise the technology.} (G1, P1)''.

\paragraph{Theme 3: Feedback on virtual focus group system.}

The third theme encapsulates certain system issues encountered during the use of Focus Agent.
A concern raised by some participants was the insufficient time allocated for responding to questions, resulting in interruptions by the agent. As articulated by one participant, ``\textit{There were instances where we were attempting to respond to a question or had just commenced our response when the moderator interrupted us and swiftly moved on to the next question}'' (G3, P3). Furthermore, two participants recommended the incorporation of subtitles to augment their understanding of the questions posed.

\section{Discussion}

In this discussion, we address the RQs through our findings and expand on the underlying reasons informed by our analysis.

\subsection{RQ1: To what extent do the opinions generated by a LLM align with those of human participants in focus group?}

The content analysis of the focus group discussions revealed that opinions generated by AI tend to encompass a wide array of human perspectives within the designated topic. Nevertheless, these AI-generated opinions often reflected more common viewpoints, demonstrating a lack of the uniqueness commonly found in human responses. A possible explanation is that, unlike human participants, who dynamically build upon previous contributions and enrich discussions with personal experiences, AI responses largely appeared as potentially plausible experiences that might happen to people. 

This observation suggests that LLMs could serve as a tool for researchers aiming to streamline the focus group process with human participants. By deploying a Focus Agent, researchers could initially gather a broad spectrum of common opinions on a specific topic, thereby setting a foundational understanding of the expected participant responses. This could further assist in refining the focus group's questions and topics, making the discussion more targeted and efficient. Therefore, fewer human focus group sessions may be required to confirm the AI-generated content and identify novel insights from participants, optimising the research process while still uncovering the unique, creative perspectives that only human participants can provide. However, human participants are still necessary for current focus groups to make sure the data is reliable.

\subsection{RQ2: To what extent is a LLM effective in performing the duties of a moderator in focus group discussions?}

During the focus group simulations, LLMs demonstrated sufficient knowledge to facilitate the group and engage with AI participants effectively. Feedback from the meta focus group indicated that human participants acknowledged the AI moderator's capability to support the discussion, albeit perceiving it more as a tool rather than a sentient interlocutor. This perception was attributed to the AI moderator's lack of apparent intelligence in interactions, such as overlooking participant requests, posing repetitive questions, or failing to grasp the hints behind conversations.

The primary challenges are rooted in the inherent limitations of LLMs in navigating multi-person dialogues. For LLMs to respond appropriately, they must comprehend inputs from human participants, reason through the conversation's context, and formulate accurate responses. While their reasoning capacity seemed adequate during simulations, issues predominantly arose in understanding and response generation phases. Existing research has begun to address LLMs' comprehension issues when assisting humans, yet their effectiveness in multi-participant discussions remains constrained~\cite{dong2023towards}. From an understanding standpoint, discussions among human participants often involve colloquial language and incomplete sentences, differing markedly from the more structured exchanges with AI, leading to the AI moderator's difficulties in recognising whether questions had been answered. Consequently, the AI moderator might repetitively address the same points rather than progressing the discussion. Additionally, challenges in generating aligned and unbiased content persist within LLM outputs~\cite{wang2023aligning, taubenfeld2024systematic}. Although not the primary focus of our study, participants noted issues such as deviation from guidelines or biases in the discussion (see \autoref{sec:meta_focus_group}), underscoring the LLMs' limitations in mimicking human conversational norms accurately.

Given these observations, we advise against deploying the Focus Agent as the sole moderator in focus group discussions due to the current inadequacies in human-AI communication. Instead, the AI-generated summaries and questions could be utilised by human moderators to streamline the discussion flow and address specific topics. For more in-depth discussions, the presence of a human moderator is essential to ensure a positive user experience and foster the generation of innovative insights, highlighting the complementary roles of AI and human moderators in enhancing the efficacy of focus groups.

\subsection{Improvement of Focus Agent}

Based on the process of the user study and the insights gathered during the meta focus group, several areas for enhancing the structure and functionality of the Focus Agent have been identified:

1. \textit{Design of thought chain:} Although the thought chain in our work shows enough ability to facilitate the focus group discussion, to be able to facilitate a deeper topic during discussion a more complex design is required, for example one such as the tree of thoughts~\cite{yao2023tree}.

2. \textit{Subtitles for the Focus Agent's speech:} Participants suggested that the speech of the Focus Agent might be too long for them to be able to comprehend a question in its entirety. In this case, subtitles would be a useful help. 

3. \textit{Time schedule of Focus Agent:} the Time allocation was a pre-determined time duration. However, the time should be allocated  according to the participants' engagement in the current discussion. In this case, the Focus Agent should make dynamic time allocations based on the flow of the discussion.

\section{Limitation and Future Work}

Our investigation underscores several limitations that pave the way for future research directions.

First, the current iteration of the Focus Agent is limited to text-based interactions, differing significantly from the multi-modal nature of human moderation. Human moderators use non-verbal cues and physical context to tailor their approach, which text-only agents cannot replicate. This limitation is particularly challenging in settings involving tactile or visual elements. However, advancements in sophisticated LLMs like GPT-4, which understand multi-modal data~\cite{openai2023gpt}, could evolve the Focus Agent into a more versatile, multi-modal platform that closely simulates human discussions.

Second, our study centres on LLM application within focus groups, overlooking broader quantitative and qualitative research methodologies. Prior studies have used LLMs to generate reviews or comments~\cite{liang2023can, chuang2023simulating}, noting that LLM-generated opinions may lack human creativity~\cite{bender2021dangers}. Ensuring the validity of these insights requires extensive empirical validation.

Lastly, our analysis highlights the difficulties LLMs face in multi-participant discussions. While there is some research on one-on-one dialogues and all-AI discussions~\cite{abbasiantaeb2024let}, studies on mixed human-AI communication in group settings are scarce. The ability of LLMs to engage in multi-human conversations is crucial for advancing human-AI interaction. Future research should explore how human participants adjust their communication strategies in the presence of AI, aiming to optimise these interactions for better collaborative outcomes.

\section{Conclusion}
Our research introduced the Focus Agent, a novel AI simulation system developed to simulate focus group discussions through the dialogue of AI agents. This system aims to gather insights akin to those derived from traditional focus groups, leveraging the capabilities of AI participants to generate discussions on designated topics. To assess the degree of alignment between the viewpoints expressed by AI and human participants, we ran a user study that employed an AI moderator to facilitate discussions among human participants. Our analysis uncovered that the Focus Agent includes opinions that similar to those of human participants. Additionally, we studied human participants’ perceptions of the AI moderator and found that while the AI could fulfil the functional role of a moderator, there remained some differences in the interaction experience compared to engagement with human moderators. We examined the underlying reasons and identified specific areas within the large language model’s capabilities that require further enhancement.

\bibliographystyle{ACM-Reference-Format}
\bibliography{main}

\end{document}